*Accepted to the Low Temperature Physics (Invited paper)*

**A colossal dielectric response of $Bi_{1-x}Sm_xFeO_3$ nanopowders**

Vladyslav O. Kolupaiev[1*], Olexander S. Pylypchuk[1], Vladimir N. Poroshin[1], Denis O. Stetsenko[1], Ihor V. Fesych[2], Lesya D. Demchenko[3,4], Eugene A. Eliseev[5], Victor V. Vainberg[1†], and Anna N. Morozovska[1‡]

[1] *Institute of Physics, National Academy of Sciences of Ukraine, 46 Nauky Ave, Kyiv, 03028, Ukraine*

[2] *Taras Shevchenko National University of Kyiv, 01030 Kyiv, Ukraine*

[3] *Stockholm University, Department of Chemistry, Sweden*

[4] *University of Ukraine "Igor Sikorsky Kyiv Polytechnic Institute", 37, Beresteisky Avenue, Kyiv, Ukraine, 03056*

[5] *Frantsevich Institute for Problems in Materials Science of the National Academy of Sciences of Ukraine, 3, str. Omeliana Pritsaka, 03142 Kyiv, Ukraine*

**ABSTRACT**

The dielectric permittivity of the pressed powder samples of $Bi_{1-x}Sm_xFeO_3$, with Sm content "x" varying in the range 0 – 0.2, has been investigated in the temperature range from 20 to 400°C and the frequency range from 100 Hz to 100 kHz. We have shown that the Sm content impacts significantly the real and imaginary parts of effective dielectric permittivity, which have expanded diffuse maxima with a colossal magnitude up to $10^5$ (for the real part) and up to $10^8$ (for the imaginary one) at temperatures 300 – 400 K. Analysis of experimental data carried has shown that both the real and imaginary parts of effective dielectric permittivity may be comprehensively explained by considering a complex interplay of a diffuse ferroelectric-paraelectric phase transition and the Maxwell-Wagner-Sillars effects, which emerge from the formation of spatial charges at interfaces between nanograins and at the ferroelectric nanoparticle-air interface. Processing of experimental data for the real and imaginary parts of the effective dielectric permittivity within effective medium approach allows us to separate and analyze the colossal dielectric response of the nanoparticles itself. The main trends followed from experiments are supported by the theoretically simulated dependences, which reveal correlations between the temperature behavior of dielectric properties and phase state of the $Bi_{1-x}Sm_xFeO_3$ nanoparticles.

* corresponding author, e-mail: kolupaiev.v.o@gmail.com
† corresponding author, e-mail: viktor.vainberg@gmail.com
‡ corresponding author, e-mail: anna.n.morozovska@gmail.com

V. O. Kolupaiev, ORCID: https://orcid.org/0009-0006-5957-6735, O. S. Pylypchuk, ORCID: https://orcid.org/0000-0003-0136-0799, V. N. Poroshin, ORCID: https://orcid.org/0000-0001-8217-3949, Denis O. Stetsenko, ORCID: 0009-0002-9141-0939, Lesya D. Demchenko, ORCID: https://orcid.org/0000-0002-2325-1940 , Ihor V. Fesych, ORCID: https://orcid.org/0000-0002-4814-5642, E. A. Eliseev, ORCID: https://orcid.org/0000-0001-8124-8857, V. V. Vainberg ,ORCID: https://orcid.org/0000-0002-9840-8033, A. N. Morozovska, ORCID: https://orcid.org/0000-0002-8505-458X

## 1. INTRODUCTION

The Bi based multiferroics $BiFeO_3$ synthesized from nanosized particles and doped by Sm, La, Pr, Eu [1, 2, 3] or due to presence of oxygen vacancies, manifest high-performance polar, magnetic and magnetoelectric properties [4, 5], which can be manipulated beside of doping means also by variation of nanoparticle size [6, 7, 8] strain [9, 10] and defect [11, 12] engineering. It is important also that the polar properties of ferroelectric perovskites of the $ABO_3$ kind, can be changed significantly by the electrochemical reactions on their surface [13, 14]. In particular, the "chemical" polarization switching can occur in thin ferroelectric films [15, 16, 17].

The properties of the rare-earth doped $BiFeO_3$ nanoparticles [18, 19], fine-grained ceramics [20, 21], and nanocomposites [22] have been much less so far studied both theoretically and experimentally as compared to thin films. At the same time, the surface-driven electrochemical "ferro-ionic" states [23, 24], which emerge due to the surface adsorption (or desorption) of oxygen ions, can significantly improve the polar properties of $Bi_{1-x}Sm_xFeO_3$ nanoparticles [25, 26, 27].

In the recent works [25 - 27] we used the Ginzburg-Landau-Devonshire-Stephenson-Highland approach to calculate polar and dielectric properties of $Bi_{1-x}Sm_xFeO_3$ nanoparticles and to analyze their phase diagrams in dependence on the nanoparticle average size and samarium content "x". In this work we investigate experimentally and analyze theortically the colossal dielectric permittivity of the pressed $Bi_{1-x}Sm_xFeO_3$ nanopowders in a wide frequency and temperature ranges. The observed features are explained by the complex interplay between a possible diffuse ferroelectric-paraelectric phase transition and the Maxwell-Wagner-Sillars (MWS) effects, which emerge from the formation of spatial charges at interfaces between different materials, and in the ferroelectric nanoparticle-air interface.

As it was shown in Refs. [21, 28, 29], the MWS-type effective dielectric permittivity reaches colossal values (greater than $10^4$ – $10^5$) at low frequencies. Mesoscopic inhomogeneities in electrical conductivity (mainly between grains/particles and their boundaries) give rise to interfacial charge accumulation, producing MWS-type polarization and internal barrier layer capacitance (IBLC), resulting in the experimentally observed colossal effective permittivity in nanograined ferroelectric ceramics, nanopowders, and/or nanocomposites [30, 31, 32]. Additional contributions to the colossal permittivity may arise from inhomogeneous layers between the electrodes and the sample, known as surface barrier layer capacitance (SBLC) [28-31]. Both the IBLC and SBLC effects can produce a colossal dielectric permittivity ($>10^4$) of $Bi_{1-x}Sm_xFeO_3$ fine-grained ceramics at low frequencies (1 kHz or less), but the permittivity decreases to $10^2$ (or less) with increase in frequency above 1 MHz (see Fig. 1 in Ref. [21]). These interfacial polarization mechanisms can be described by the effective medium approach (EMA) [33], as discussed in Refs. [34, 35], and in the works of Petzelt et al. [36] and Richetsky et al. [37]. Generally speaking, for the systems, where the sharp enhancement of the dielectric permittivity due to the diffuse (or sharp) ferroelectric-paraelectric transition dominates, the dielectric response is weakly frequency-independent up to the MHz frequency range.

In this work, we have studied the character of relaxation in processes in the pressed $Bi_{1-x}Sm_xFeO_3$ nanopowder samples. The Sm content "x" varies in the range 0 – 0.2. For this purpose, we performed analysis of the measured dependences of the real and imaginary parts of the effective dielectric permittivity in the temperature range 20 – 400 $^{o}$C and frequency range 4 Hz – 500 kHz. To reveal general experimental trends inherent to low and high frequencies, the measured temperature dependences are presented and analyzed for two fixed frequencies – 100 Hz and 100 kHz. The detailed frequency dependences were measured from 4 Hz to 500 kHz at room temperature. Theoretical analysis has been performed within the framework of effective medium approach (EMA). This allows us to separate and analyze the colossal dielectric response of the nanoparticles themselves. The main trends of experimental observations have been theoretically simulated, which allows us to establish correlations between the temperature behavior of dielectric properties, domain structure morphology and phase state of the $Bi_{1-x}Sm_xFeO_3$ nanoparticles.

## 2. MATERIALS AND METHODS

A series of nanopowder samples of $Bi_{1-x}Sm_xFeO_3$, where the Sm content "x" varies from 0 to 0.2, was prepared using the solution combustion method and further calcination for 5 hours at 750 °C to minimize the presence of residual water and hydroxyl groups. Preparation details are given in Ref. [27]. The Sm content is x = 0, 0.05, 0.1, 0.15, 0.2.

According to the X-ray diffraction (XRD) data [27], the $BiFeO_3$ sample contains 73 % of the long-range ordered rhombohedral (*R*3*c*) $BiFeO_3$ phase, 16% of the orthorhombic (*Pbam*) $Bi_2Fe_4O_9$ phase, and 11% of the cubic (*I*23) $Bi_{25}FeO_{40}$ phase. The Sm-doping very strongly increases the phase purity of the nanopowders. So the sample with nominal Sm content x = 0.05 contains 97% of $Bi_{0.95}Sm_{0.05}FeO_3$ in the *R*3c phase, 2% of $Bi_2Fe_4O_9$, and 1% of $Bi_{25}FeO_{40}$. The nanopowders with nominal Sm content x = 0.10 contain 99% of $Bi_{0.9}Sm_{0.1}FeO_3$ in the *R*3c phase and 1% of $Bi_{25}FeO_{40}$. The nanopowders with nominal Sm content x = 0.15 contains 7% and 92% of $Bi_{0.85}Sm_{0.15}FeO_3$ in the polar *R3c* and orthorhombic *Pbnm* phases, respectively, and 1% of $Bi_{25}FeO_{40}$. The nanopowders with nominal Sm content x = 0.20 contains 99% of $Bi_{0.8}Sm_{0.2}FeO_3$ in the *Pbnm* phase and 1% of $Bi_{25}FeO_{40}$.

To measure the temperature dependences of the dielectric response, the $Bi_{1-x}Sm_xFeO_3$ nanopowders were pressed in the polytetrafluoroethylene (PTFE) cells between two metallic plungers, which serve as electric contacts (see **Fig. 1**). The sample in the cell has a disk shape with the 4 mm diameter and 0.2 mm thickness. The pressure to keep the samples resistance within measurable limits was about 2.5 MPa. The higher pressure applied to the samples does not result in noticeable changes in their electrophysical characteristics, and the measured values do not change after removing pressure. The PTFE cell with plungers, providing sample compression, was placed inside the holder connected to the RLC meter *via* the coaxial cable. The RLC meter UNI-T UT612 was used to measure the capacitance of the cells in study of temperature dependences of the dielectric permittivity at fixed frequencies of 100 Hz and 100 kHz. The samples heating was performed by the wire-wound nichrome resistive heater connected to the controlled voltage supply. The RLC-meter LCX200 ROHDE & SCHWARZ was used for the measurements of capacitance vs frequency in the range 4 Hz – 500 kHz at the room temperature.

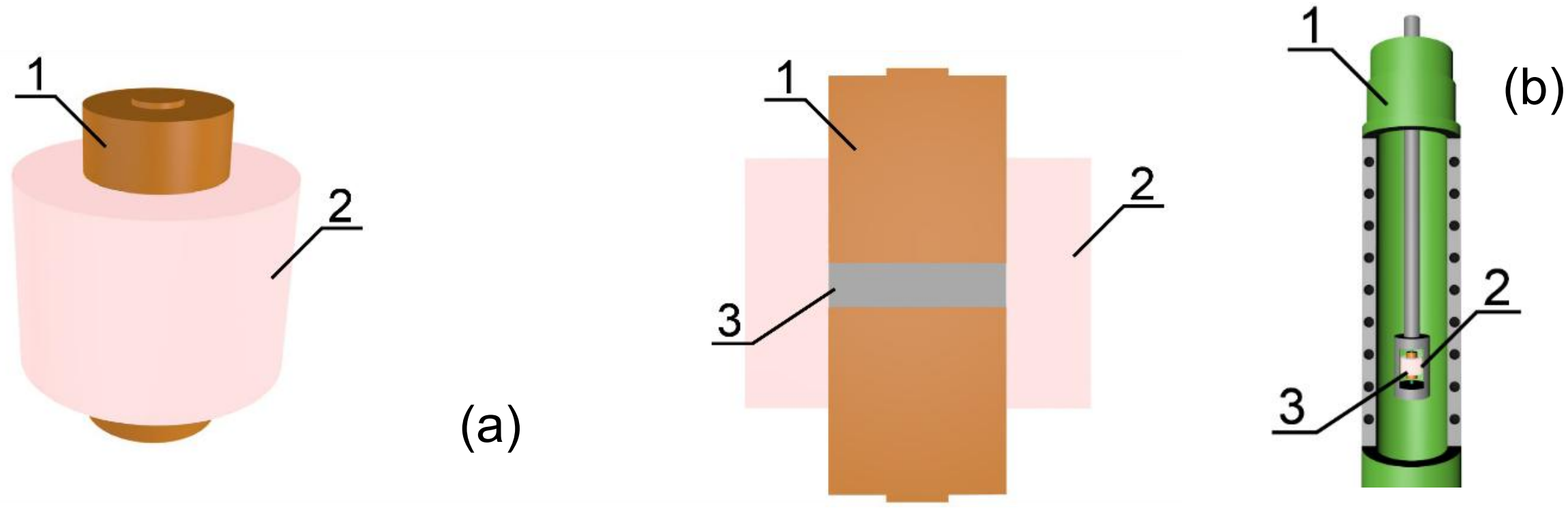

**Figure 1.** (a) Schematic diagram of the PTFE cell with nanopowder: (1) – brass contacts; (2) – PTFE tube; (3) – research sample. (b) Schematic diagram of the heating chamber: (1) – resistive heater; (2) – holder; (3) – PTFE cell.

Typical transmission electron microscopy (TEM) images of the $Bi_{1-x}Sm_xFeO_3$ nanoparticles are shown in **Figs. 2(b) – 2(f).** As seen from the images, the spread of nanoparticle sizes is large enough; it varies from 50 nm to 500 nm. Particles tend to form large agglomerates (see e.g., **Fig. 2(b)**). The shape of individual nanoparticles is irregular (see e.g., **Fig. 2(c)**) and will be substituted by "effective" spheres in theoretical analysis.

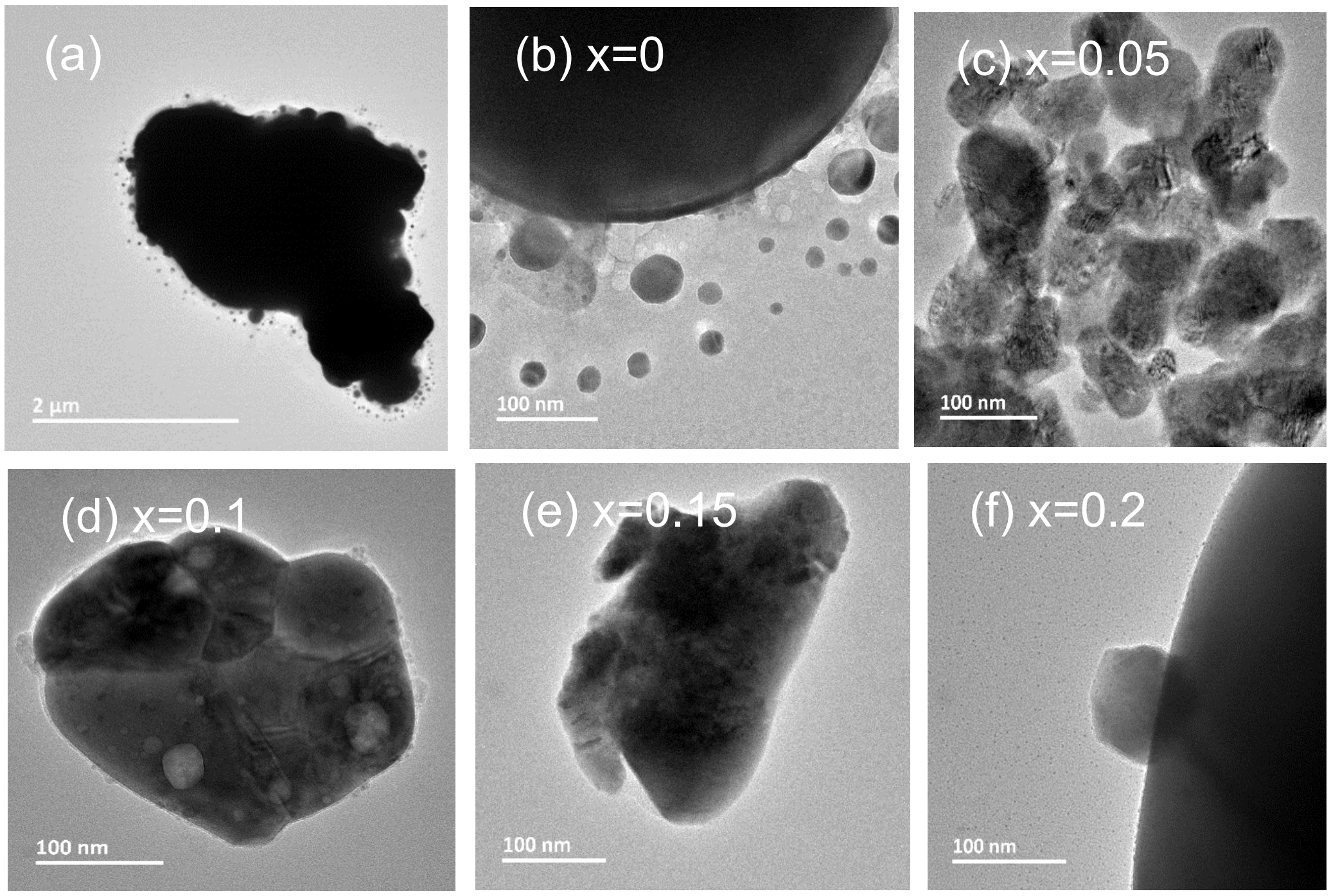

**Figure 2. (a)** Typical low-resolution TEM image of the $BiFeO_3$ nanopowder. **(b)–(f)** Typical higher-resolution TEM images of the $Bi_{1-x}Sm_xFeO_3$ ($0 \leq x \leq 0.2$) nanoparticles.

## 3. RESULTS AND DISCUSSION

### A. Experimental results and discussion

The temperature dependences of the effective dielectric permittivity of the pressed $Bi_{1-x}Sm_xFeO_3$ nanopowders were measured in the frequency range from 100 Hz to 100 kHz and in the temperature range from 20 to 400 $^{o}$C.

The dependences of the real and imaginary parts of the dielectric permittivity measured for the samples with all contents of Sm at 100 Hz and 100 kHz, are shown by solid symbols in **Fig. 3** and **4**, respectively. The empty symbols on these curves correspond to values obtained with correction for the tight packing of the nanograins as it is described below in Section B, see Eq.(4). It is seen that the initial and corrected values are very close to each other in the logarithmic scale.

It is seen that the value of both the real and imaginary parts of dielectric permittivity, substantially and non-monotonously depends on the Sm content. At that, qualitatively, in general, the dependences have similar behavior for all Sm content "x" and consist of two temperature regions. In the first range of 20 – 300ºC and at frequency $f$ =100 Hz, real and imaginary parts of the effective dielectric permittivity of all samples very weakly depends on temperature manifesting only very small increase in magnitude with increase in temperature. At 100 kHz the upper limit shifts downward approximately to 250 ºC. In the second temperature region of 300 – 400ºC at 100 Hz and 250 – 400 at 100 kHz, the real and imaginary parts of the effective dielectric permittivity demonstrate a strong increase with increase in temperature, with the tendency to form a maximum at the highest temperature. Also, in general, the magnitude of permittivity at 100 kHz is at least by one order of magnitude less.

Also, one may note that the permittivity vs temperature curves reveal an expanded diffuse like maximum, especially for the imaginary part. This is usually a characteristic feature of nanograin powders and depends both on the particle size, spread of their size and contribution of ion conduction via the interface space.

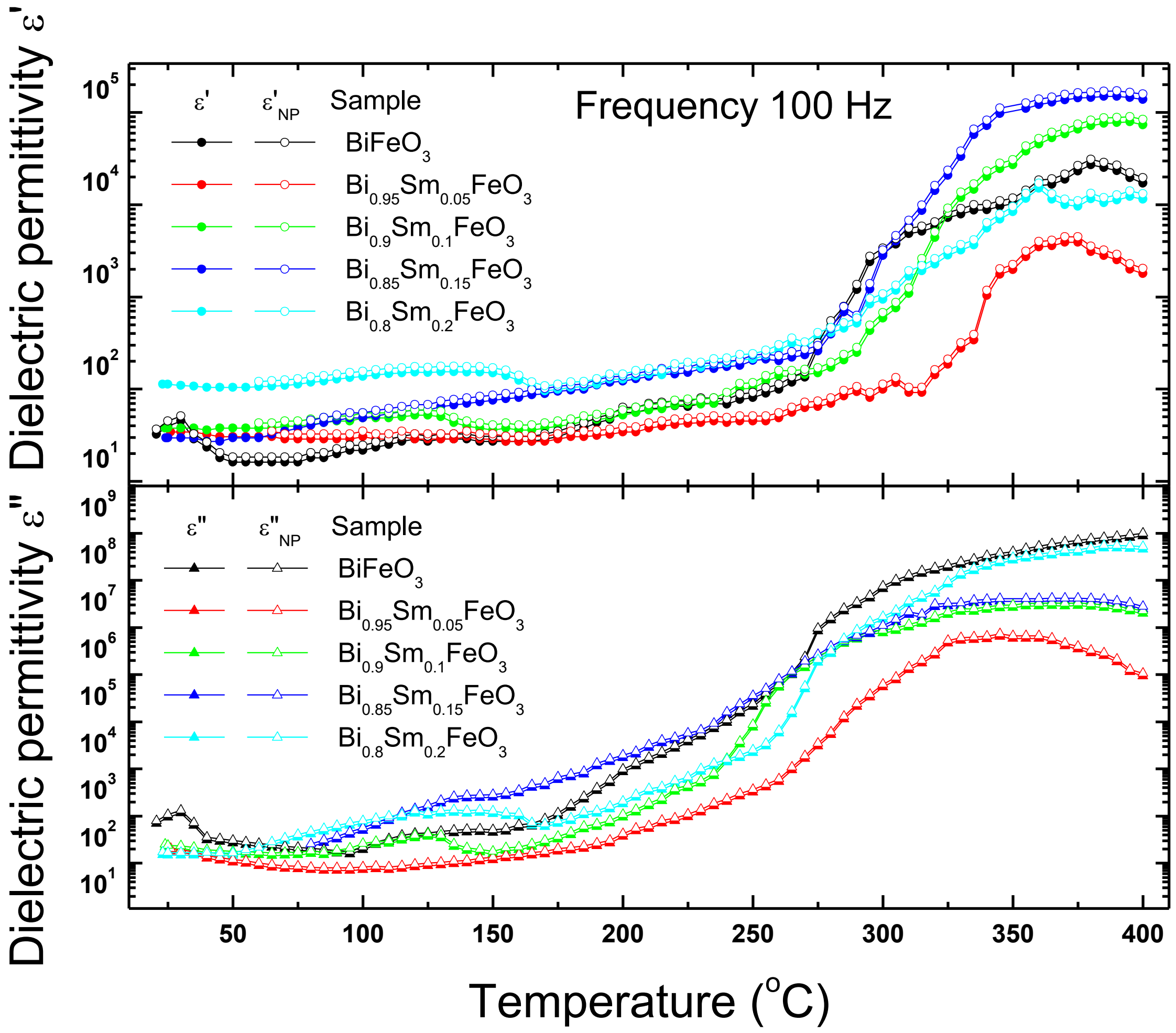

**Figure 3.** Temperature dependencies of the real $\varepsilon'$ (circles) and the imaginary $\varepsilon''$ (triangles) parts of the dielectric permittivity of the pressed $Bi_{1-x}Sm_xFeO_3$ nanopowders at 100 Hz frequency. Filled symbols are experimental results, empty symbols are recalculated according to Eq. (4).

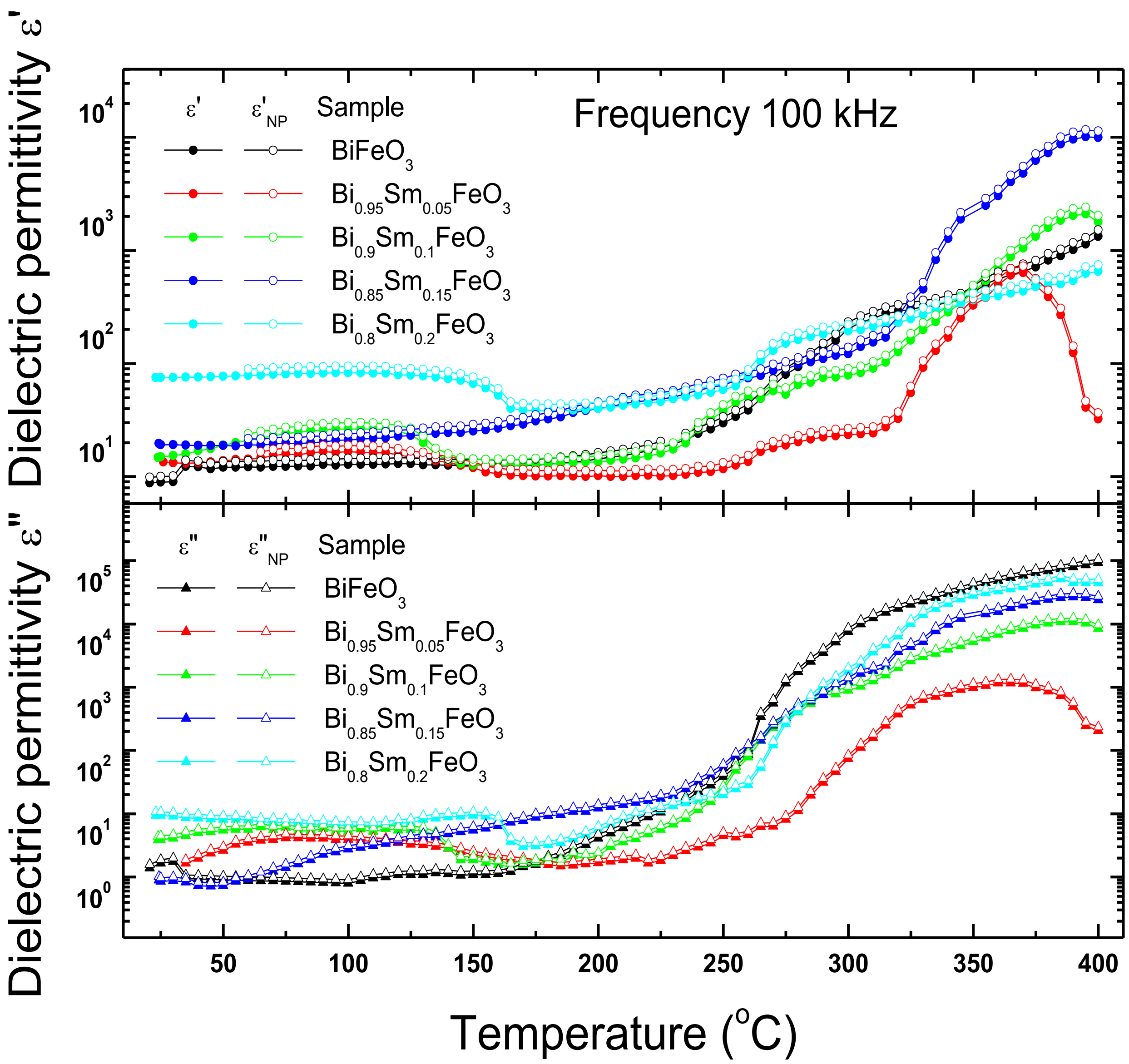

**Figure 4.** Temperature dependencies of the real $\varepsilon'$ (circles) and the imaginary $\varepsilon''$ (triangles) parts of the dielectric permittivity of the pressed $Bi_{1-x}Sm_xFeO_3$ nanopowders at 100 kHz frequency. Filled symbols are experimental results, empty symbols are recalculated according to Eq. (4).

**Figure 5** shows how the temperature of the maximum of the real and imaginary parts of dielectric permittivity varies at 100 Hz and 100 kHz, respectively. It is seen that addition of Sm at first decrease permittivity and forms a minimum. At further increase of the Sm content the permittivity returns approximately to the initial magnitude and varies differently at different Sm content. Note that at small content of Sm (less than 0.15) the behavior of the real and imaginary parts is similar, while at higher content they show tendency to change in the opposite directions. One can also note that at 100 kHz as compared to 100 Hz at the higher Sm content both the real and imaginary parts change direction of the ε vs the Sm content variation onto the opposite one. For comparison, the dependence of the dielectric permittivity vs the Sm content, measured at 4 Hz and room temperature, is shown by olive curve. It is seen that the dependence is a nonmonotonic function of "x" with a local maximum at Sm content $x \approx 0.05$ and local minimum at $x \approx 0.15$. These features, whose x-position almost coincides with the x-position of features on other dependences shown in the figure, may be considered as indication of the Sm-doping impact.

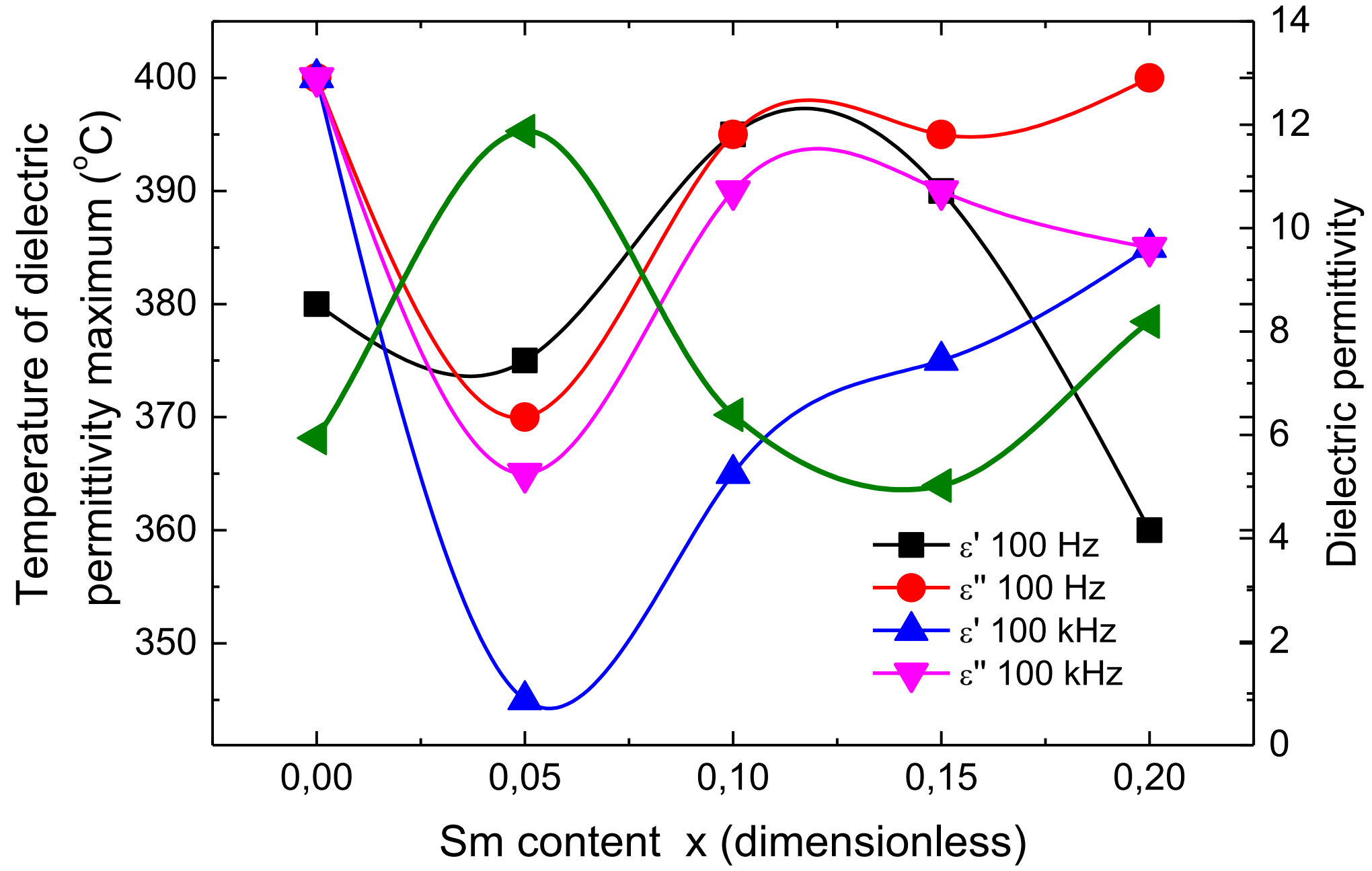

**Figure 5**. Variation of the temperature of the dielectric permittivity maximum for the real (black rectangles, blue triangles) and imaginary (red circles, magenta triangles) parts at 100 Hz and 100 kHz, respectively. The dependence of the $Bi_{1-x}Sm_xFeO_3$ nanopowders dielectric permittivity vs the Sm content, measured at 4 Hz and room temperature, is shown by olive curve.

To judge qualitatively on the spread of the maximum temperature and width of the dielectric response, and respectively on spread of physical properties related to the temperature intervals of relaxation processes in the studied samples with different Sm content, in **Fig. 6** there is shown comparison of temperature ranges widths for the range of the slow growth or almost constant dielectric permittivity vs temperature and the range of the sharp and strong growth. The panels (**a**) and (**b**) correspond to the real and imaginary parts of the complex dielectric permittivity at 100 Hz, respectively. The temperature ranges in full detail both at $f$ =100 Hz and 100 kHz are given in tables in **Appendix**.

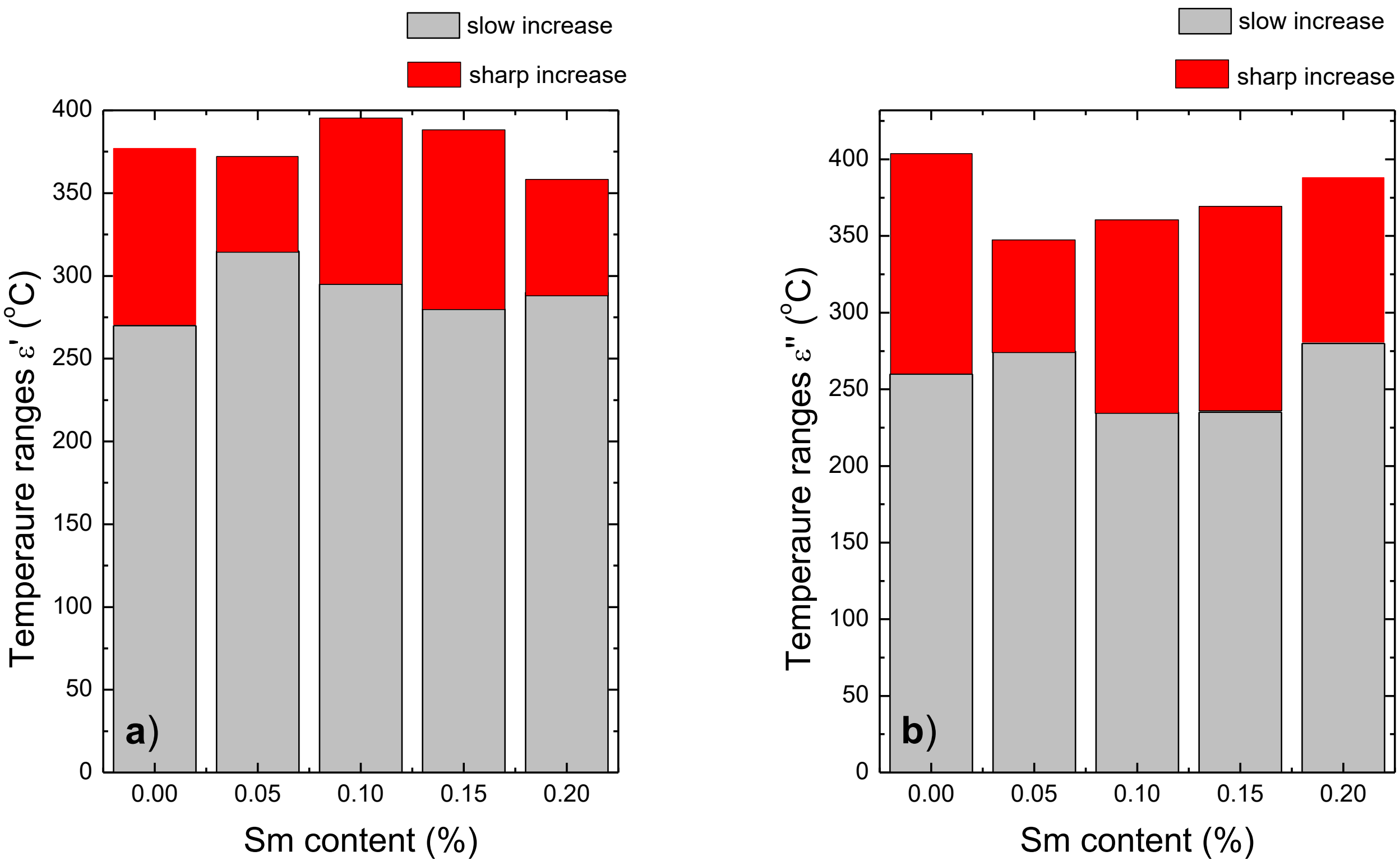

**Figure 6.** Comparison of temperature ranges of the slow and sharp growth of the real **(a)** and imaginary **(b)** parts of the $Bi_{1-x}Sm_xFeO_3$ nanopowders dielectric permittivity vs Sm content.

One may note that the spread of the width of the sharp growth range is really large, that gives evidence on structural non-homogeneity which impacts stronger on the imaginary part of permittivity which experiences more impact of the interface charge transfer mechanisms. The least spread, as seen from **Fig. 6**, is observed for the sample with 5% of the Sm content which demonstrates also the lowest temperature for the maximum of the real part of permittivity. Such picture may be considered as evidence of variation temperature boundaries of the ranges with high growth of permittivity between different agglomerates in the whole mixture caused by structural properties such as homogeneity and defects depending on size and stoichiometry.

Finally, let us analyze the Cole-Cole diagram for three characteristic samples under study with different Sm content: x = 0, x = 5 % and x = 10 % (see **Fig. 7**). The diagram shows that with the addition of Sm content, there is a systematic shift of the obtained dependences towards higher values. It is seen that all curves consist of an arc, which may be classified as a strongly deformed "semicircle" in the high-frequency range, and a sharp up-directed increasing branch in the low-frequency range. The deformed semicircles, which correspond to higher frequences, can be attributed to the permittivity-losses relation characteristic to less conductive regions, such as ferroelectric grain cores and/or dielectric intergrain air spaces, whose dielectric response corresponds to the ferroelectric capacitance response prevailing at high frequencies. The sharp increase of conductivity, which appears at low frequences, can be attributed to the release/relaxation of sluggish screening charges located in the grain shells. The low-frequency relaxation of the screening charges leading to the formation of interfacial polarization belongs to the Maxwell-Wagner type effects. The proposed interpretation of the Cole-Cole plot is consistent with the theory of the IBLC and SBLC effects [28-31] as well as with experimental observations of the dielectric response of polycrystalline samarium bismuth ferrite ceramics [21]. According to classical EMA describing IBLC and SBLC effects [36, 37], the relative contribution of less conductive ferroelectric grain cores and/or dielectric intergrain air spaces is significant at higher frequencies, while the contribution of screening charges located in the grain shells becomes significant at low frequencies. Thus, the Cole-Cole plot gives us additional evidence of the mixed multi-component character of relaxation processes, which emerge in the spatially-inhomogeneous semiconductor-dielectric system "grain core – shell – surrounding". As a rule, Sm-doping decreases the conductivity of a homogeneous single-crystalline bismuth ferrite (see Refs. [1, 5] and refs therein), but it appears not the case for the spatially inhomogeneous $Bi_{1-x}Sm_xFeO_3$ nanopowders.

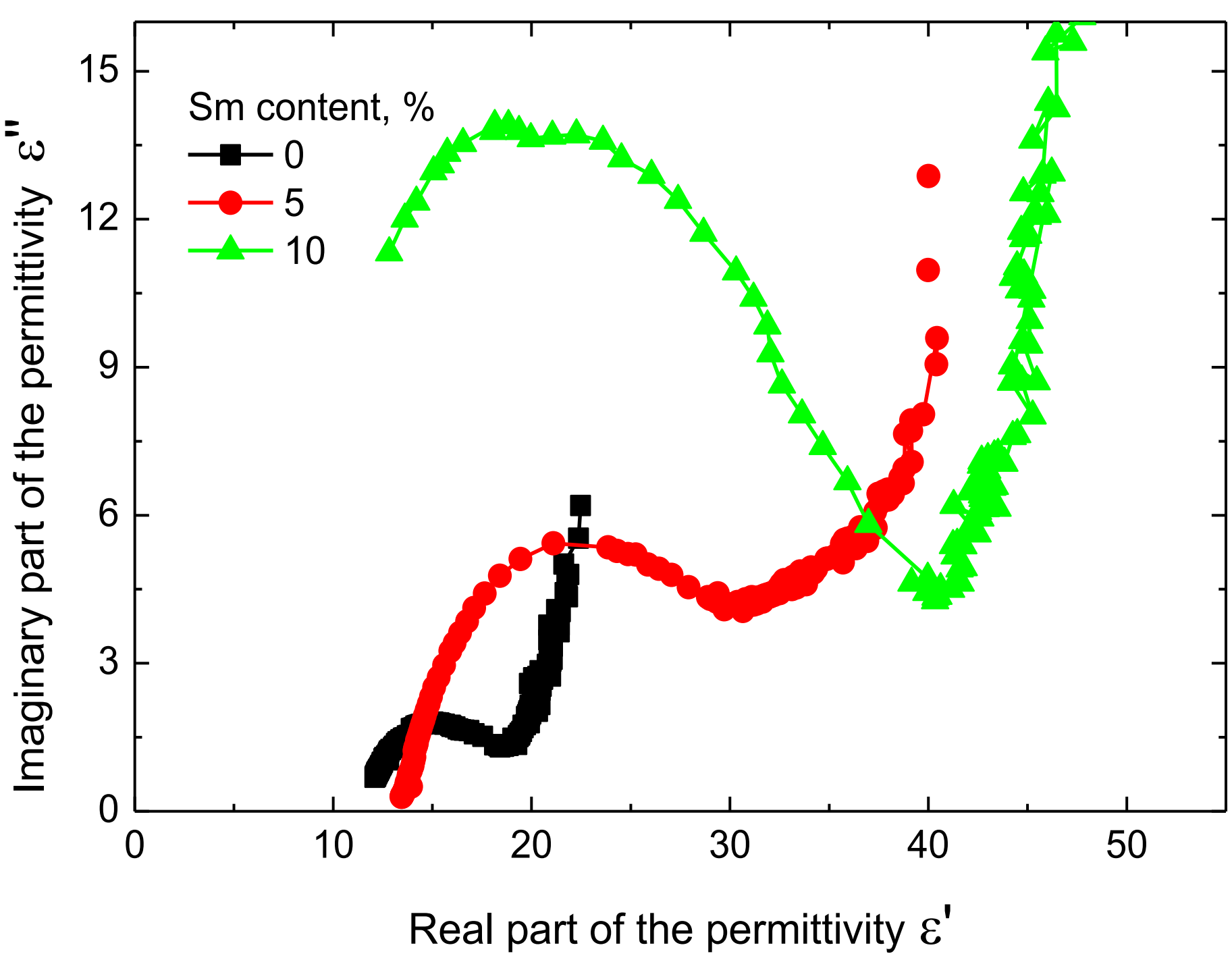

**Figure 7**. The Cole-Cole diagrams for the pressed $Bi_{1-x}Sm_xFeO_3$ nanopowder samples with Sm content x = 0 (black symbols) x = 5 % (red symbols) and x = 10 % (blue symbols) measured at 30ºC.

### B. Processing of experimental results

A possible synergy of the IBLC and SBLC effects and dipolar polarization can be analyzed within the EMA, which yields an algebraic equation for determining the complex dielectric permittivity $\varepsilon_{eff}^*$ of the effective medium [33]:

$$(1-\mu)\frac{\varepsilon_{eff}^*-\varepsilon_s^*}{(1-\eta_{NP})\varepsilon_{eff}^*+\eta_{NP}\,\varepsilon_s^*}+\mu\frac{\varepsilon_{eff}^*-\varepsilon_{NP}^*}{(1-\eta_{NP})\varepsilon_{eff}^*+\eta_{NP}\,\varepsilon_{NP}^*}=0. \quad (1)$$

Here, $\varepsilon_s^*$ is the complex relative permittivity of the dielectric surrounding "*s*", at that $\mathrm{Re}[\varepsilon_s^*] \gg \mathrm{Im}[\varepsilon_s^*]$ for air. The values $\mu$ and $1-\mu$ are relative volume fractions of the components "*NP*" and "*s*", respectively. The maximal value $\mu_{max}$ is limited by the dense packing of the particles. In particular, $\mu_{max} \approx 0.75$ for densely packed nanospheres of a uniform radius. $\varepsilon_{NP}^*$ is the complex relative permittivity of the polarized nanoparticles "*NP*", which are assumed to be monodisperse and uniformly polarized. The real part of $\varepsilon_{NP}^*$ corresponds to the dielectric (polarization) response, and the imaginary part of $\varepsilon_{NP}^*$ accounts for dielectric losses due to finite ionic conductivity, these losses give rise to MWS type and SBLC effects. The function $n_{NP}$ is the depolarization field factor of the polarized nanoparticles, which is determined by their shape and polarization orientation; $0 \leq n_{NP} \leq 1$. Analytical expressions for $n_{NP}$ exist for uniformly polarized nanoparticles of ellipsoidal shape including nanowires, nanospheres, and nanodisks [38]. Free charges, existing in the nanoparticle shell, screen the depolarization field inside the particle, and corresponding decrease of "effective" depolarization factor $\eta_{eff}$ can become much smaller.

Since the nanoparticles are not monodispersed and the spread of their sizes are large in the considered case (see the TEM images in **Fig. 2)**, the physical sense of $\varepsilon_{NP}^*$ and $n_{NP}$ should be discussed, specified and quantified. The shape of individual nanoparticles is irregular (see e.g., **Fig. 2(c)**) and will be substituted by "effective" spheres in theoretical analysis. As for effective depolarization factor, we consider two limiting cases, $\eta_{NP}=1/3$ (which corresponds to spatially isolated and weakly screened spherical nanoparticles) and $\eta_{NP}=0$ (which corresponds to the well-screened nanoparticles of arbitrary shape or/and to the case when they form percolation clusters between the electrodes). However, since the studied $Sm_xBi_{1-x}FeO_3$ nanopowders are characterized by a wide distribution of sizes (see **Fig. 2**), the question about the influence of finite size effects on the phase state of individual nanoparticles should be studied to corroborate the validity of $\varepsilon_{NP}^*$ introduction.

For the theoretical description of the influence of size effects on the phase transitions in the $Sm_xBi_{1-x}FeO_3$ nanoparticles we modified the Landau-Ginsburg-Devonshire-Stephenson-Highland (LGDSH) approach and the model of four sublattices (FSM) (see Refs. [39, 40, 41] for details). Within the modified LGDSH+FSM approach, the $Sm_xBi_{1-x}FeO_3$ are characterized by the sublattice-sensitive dipolar polarization vectors $\vec{P}^{(i)}$ and the axial antiferrodistortive (AFD) tilt pseudovectors $\vec{\Phi}^{(i)}$ (*i*=1-4) [41]. The ferroelectric polar and antiferroelectric order parameters are $\vec{P}=\frac{1}{2}\left(\vec{P}^{(1)}+\vec{P}^{(2)}+\vec{P}^{(3)}+\vec{P}^{(4)}\right)$ and $\vec{A}=\frac{1}{2}\left(\vec{P}^{(1)}-\vec{P}^{(2)}+\vec{P}^{(3)}-\vec{P}^{(4)}\right)$, respectively. The spatial orientation and magnitude of $\vec{P}$ and $\vec{A}$ determines the phase. The FE phase corresponds to $P\neq 0$ and $A=0$; the FEI phase corresponds to $P\neq 0$ and $A\neq 0$, and the AFE phase corresponds to $A\neq 0$ and $P=0$. The nonpolar phase corresponds to $A=0$ and $P=0$.

Within the modified LGDSH+FSM approach the description of the $Sm_xBi_{1-x}FeO_3$ PE phase diagram is reduced to the thermodynamic analyses of the Landau-type free energy with coefficients renormalized by the oxygen tilt subsystem [40, 41]. The renormalized LGD-part of the free energy, which determines the phase state, is [41]:

$$G_{LGD}=\frac{1}{2}\alpha(T,x)\vec{P}^2+\frac{1}{2}\eta(T,x)\vec{A}^2+\frac{1}{4}\left(\vec{P}^4+\vec{A}^4\right)+\frac{\delta}{2}\vec{A}^2\vec{P}^2. \qquad (2)$$

The temperature, size and elastic stress dependences of the dimensionless stiffnesses of the sublattices, $\alpha(T,x)$ and $\eta(T,x)$ are considered in Refs. [25, 40, 41]. Note that the flexoelectric-AFD coupling should be included to the LGD free energy functional, since the coupling is responsible for appearance of interfacial polarization induced by oxygen octahedral rotations at the antiphase boundaries and/or twin walls in $SrTiO_3$ [42, 43, 44], as well as it leads to the appearance of versatile spatially modulated structures in $Sm_xBi_{1-x}FeO_3$ [45, 46]. The coupling renormalizes the strength $\delta$ in Eq.(2).

For $Sm_xBi_{1-x}FeO_3$ thin films [40, 41] and nanoparticles [25], the modified LGDSH+FSM approach predicts the gradual transition from the single ferroelectric (FE) phase of rhombohedral (R) symmetry to the ferrielectric (FEI) phase of mixed rhombohedral + tetragonal (R+T) symmetry, then to the of quadruple phase (QFEI) with rhombohedral + tetragonal + orthorhombic polar + orthorhombic nonpolar (R+T+$O_p$+$O_{NP}$) symmetry, then to the binary antiferroelectric (AFE) phase of tetragonal + nonpolar orthorhombic (T+$O_{NP}$) symmetry, and eventually to the nonpolar (NP) AFD phase of orthorhombic ($O_{NP}$) symmetry with increase in Sm content from 0 to 20 % (see Refs. [40, 41] for details). Using this approach, we theoretically simulate the phase transition sequence in the 50-nm and 500-nm $Sm_xBi_{1-x}FeO_3$ nanoparticles related with the case of the Sm/Bi cation sublattice, which are schematically shown in **Fig. 8(a)** and **8(b)**, respectively**.** As one can see from comparison of these figures, the influence of size effects is negligible for 50-nm nanoparticles in the temperature range above 300 K and virtually absent for 500-nm nanoparticles. This is due to the diffuse boundary between the rhombohedral FE phase and quadruple phase QFEI, that is in fact a morphotropic boundary.

Assuming that either $\eta_{NP}\rightarrow 0$ or $\eta_{NP}=1/3$, explicit expressions for $\varepsilon^*_{NP}$ can be derived from Eq.(1). They have a relatively simple form:

$$\varepsilon_{NP}^{*} = \begin{cases} \frac{\varepsilon_{eff}^{*}}{\mu} - \varepsilon_{S}^{*}\frac{1-\mu}{\mu}, & \eta_{NP} \to 0, \\ \varepsilon_{eff}^{*}\frac{\mu\left[2\varepsilon_{eff}^{*}+\varepsilon_{S}^{*}\right]+2(1-\mu)\left(\varepsilon_{eff}^{*}-\varepsilon_{S}^{*}\right)}{\mu\left[2\varepsilon_{eff}^{*}+\varepsilon_{S}^{*}\right]-(1-\mu)\left(\varepsilon_{eff}^{*}-\varepsilon_{S}^{*}\right)}, & \eta_{NP} = \frac{1}{3}. \end{cases} \tag{3}$$

In this work we suppose that we have densely packed weakly screened ferroelectric nanoparticles situated in air (i.e. $\varepsilon_{S}^{*} = 1, \eta_{NP} = \frac{1}{3}, \mu = 0.75$) for the processing of the experimental results, so we use the second line of Eq.(2):

$$\varepsilon_{NP}^{*} = \varepsilon_{eff}^{*}\frac{0.75\left[2\varepsilon_{eff}^{*}+1\right]+0.5\left(\varepsilon_{eff}^{*}-1\right)}{0.75\left[2\varepsilon_{eff}^{*}+1\right]-0.25\left(\varepsilon_{eff}^{*}-1\right)}. \tag{4}$$

Experimental results measured at 100 Hz and 100 kHz and processed using Eq.(4), are shown by empty symbols in **Fig. 3** and **4**, respectively. It is evident that the effect of colossal dielectric response remained for both lower (100 Hz) and higher (100 kHz) frequencies.

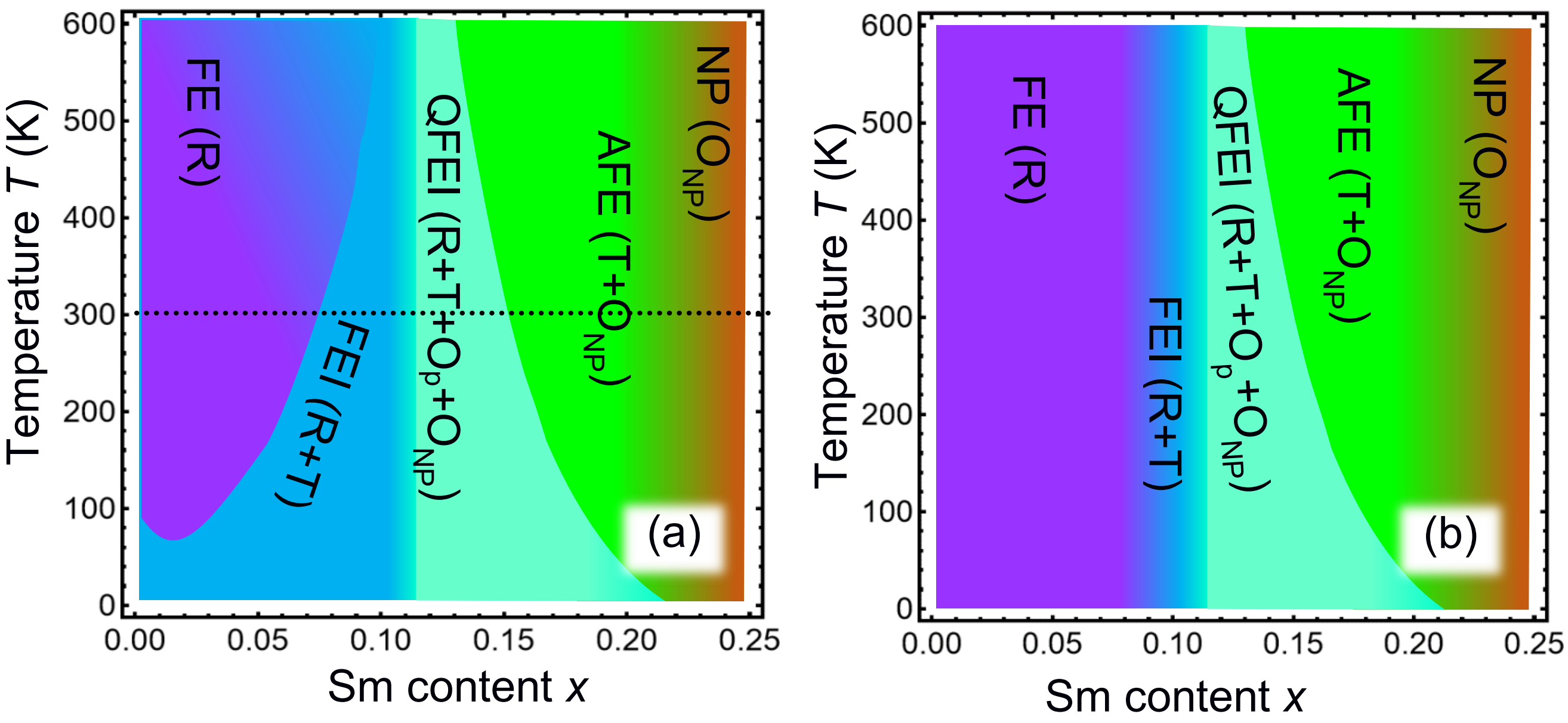

**Figure 8.** Phase diagram of the $Sm_xBi_{1-x}FeO_3$ nanoparticles with the average size 50 nm **(a)** and 500 nm **(b)** as a function of Sm content $x$ and temperature $T$ calculated for the parameters listed in Refs. [40], which correspond to the best fitting with experimental results shown in the paper.

## 4. CONCLUSIONS

We investigated experimentally and analyzed the colossal complex dielectric permittivity of pressed $Bi_{1-x}Sm_xFeO_3$ nanopowders, in the wide frequency and temperature ranges. The observed features are explained by the complex interplay between a possible diffuse ferroelectric-paraelectric phase transition and the Maxwell-Wagner-Sillars effects, which emerge from the formation of spatial charges at interfaces between different materials, as well as at the ferroelectric nanoparticle-air interface.

Processing of experimental data for the real and imaginary parts of the effective dielectric permittivity, performed within the frames of the effective medium approach, allows us to separate and analyze the colossal dielectric response of the nanoparticles itself. The main trends of experimental observations are explained also by theoretical simulation, which allows us to establish correlations between the temperature behavior of dielectric properties and phase state of the $Bi_{1-x}Sm_xFeO_3$ nanoparticles.

## APPENDIX

**Table 1.** Peculiarities of the real **(a)** and imaginary **(b)** parts of the dielectric permittivity of the pressed $Bi_{1-x}Sm_xFeO_3$ nanopowders at 100 Hz

| **(a)** Real part of the dielectric permittivity at 100 Hz | | | | |
|---|---|---|---|---|
| Sample | Maximum, °C | Slow increase, °C | Fast increase, °C | Decrease, °C |
| $BiFeO_3$ | 380 | 20.6-270 | 270-380 | 380-400 |
| $Bi_{0.95}Sm_{0.05}FeO_3$ | 375 | 26-315 | 315-375 | 375-400 |
| $Bi_{0.9}Sm_{0.1}FeO_3$ | 395 | 23.9-295 | 295-395 | 395-400 |
| $Bi_{0.85}Sm_{0.15}FeO_3$ | 390 | 24.2-280 | 280-390 | 390-400 |
| $Bi_{0.8}Sm_{0.2}FeO_3$ | 360 | 23.1-290 | 290-360 | 360-400 |
| **(b)** Imaginary part of the dielectric permittivity at 100 Hz | | | | |
| Sample | Maximum, °C | Slow increase, °C | Fast increase, °C | Decrease, °C |
| $BiFeO_3$ | 400 | 20.6-260 | 260-400 | - |
| $Bi_{0.95}Sm_{0.05}FeO_3$ | 345 | 26-275 | 275-345 | 345-400 |
| $Bi_{0.9}Sm_{0.1}FeO_3$ | 365 | 23.9-235 | 235-365 | 365-400 |
| $Bi_{0.85}Sm_{0.15}FeO_3$ | 375 | 80-235 | 235-375 | 375-400 |
| $Bi_{0.8}Sm_{0.2}FeO_3$ | 385 | 23.1-280 | 280-385 | 385-400 |

**Table 2.** Peculiarities of the real **(a)** and imaginary **(b)** parts of the dielectric permittivity of the pressed $Bi_{1-x}Sm_xFeO_3$ nanopowders at at 100 kHz

| **(a)** Real part of the dielectric permittivity at 100 kHz | | | | |
|---|---|---|---|---|
| Sample | Maximum, °C | Slow increase, °C | Fast increase, °C | Decrease, °C |
| $BiFeO_3$ | 400 | 20.6-260 | 260-400 | - |
| $Bi_{0.95}Sm_{0.05}FeO_3$ | 370 | 26-320 | 320-370 | 370-400 |
| $Bi_{0.9}Sm_{0.1}FeO_3$ | 395 | 23.9-310 | 310-395 | 395-400 |
| $Bi_{0.85}Sm_{0.15}FeO_3$ | 395 | 24.2-315 | 315-395 | 395-400 |

| $Bi_{0.8}Sm_{0.2}FeO_3$ | 400 | 165-260 | 260-400 | 23.1-165 |
| --- | --- | --- | --- | --- |
| **(b)** Imaginary part of the dielectric permittivity at 100 kHz | | | | |
| Sample | Maximum, °C | Slow increase, °C | Fast increase, °C | Decrease, °C |
| $BiFeO_3$ | 400 | 20.6-260 | 260-400 | - |
| $Bi_{0.95}Sm_{0.05}FeO_3$ | 365 | 26-280 | 280-365 | 365-400 |
| $Bi_{0.9}Sm_{0.1}FeO_3$ | 390 | 23.9-260 | 260-390 | 390-400 |
| $Bi_{0.85}Sm_{0.15}FeO_3$ | 390 | 24.2-265 | 265-390 | 390-400 |
| $Bi_{0.8}Sm_{0.2}FeO_3$ | 385 | 165-260 | 260-385 | 23.1-165; 385-400 |

**Authors' contribution.** V.O.K., O.S.P., and V.N.P. performed dielectric measurements and analyzed results. I.V.F. prepared the samples and characterized them. L.D.D. performed TEM analysis. O.S.P. and D.O.S. prepared figures. E.A.E. performed numerical modelling and prepared corresponding figures. A.N.M. formulated the theoretical problem, performed analytical calculations and analyzed results. A.N.M., V.O.K., and V.V.V. wrote the paper. V.N.P. coordinated the research.

**Acknowledgements.** The work of O.S.P., D.O.S. and A.N.M. are funded by the National Research Foundation of Ukraine (project "Manyfold-degenerated metastable states of spontaneous polarization in nanoferroics: theory, experiment and perspectives for digital nanoelectronics", grant N 2023.03/0132). V.O.K., V.V.V. and V.N.P. acknowledges the Target Program of the National Academy of Sciences of Ukraine, Project No. 5.8/25-П "Energy-saving and environmentally friendly nanoscale ferroics for the development of sensorics, nanoelectronics and spintronics". The work of E.A.E. is funded by the National Research Foundation of Ukraine (project "Silicon-compatible ferroelectric nanocomposites for electronics and sensors", grant N 2023.03/0127). L.D. acknowledges support from the Knut and Alice Wallenberg Foundation (grant no. 2018.0237) for TEM research.